# Ferroelectric Antiferromagnetic Quantum Anomalous Hall Insulator in Two-Dimensional van der Waals Materials

Yan Liang[1], Fulu Zheng[2], Thomas Frauenheim[2,3,4,*], Pei Zhao[1,*]

[1] *College of Physics and Optoelectronic Engineering, Faculty of Information Science and Engineering, Ocean University of China, Songling Road 238, Qingdao 266100, People's Republic of China;*

[2]*Bremen Center for Computational Materials Science, University of Bremen, 28359 Bremen, Germany;*

[3]*Beijing Computational Science Research Center, 100193 Beijing, People's Republic of China;*

[4]*Shenzhen JL Computational Science and Applied Research Institute, 518109 Shenzhen, People's Republic of China.*

*Corresponding author: thomas.frauenheim@bccms.uni-bremen.de; zhaopei@ouc.edu.cn

Ferroelectricity, anti-ferromagnetism (AFM) and quantum anomalous Hall effect (QAHE) are three fundamental phenomena in the field of condensed matter physics, which could enable the realization of novel devices and thus attracts great attention. Here, we show theoretical evidence that two-dimensional (2D) even-layer $MnBi_2Te_4$ allows for the simultaneous presence of intercorrelated ferroelectricity, AFM, and QAHE. Importantly, through rational van der Waals sliding, these exotic properties are strongly coupled. Such coupling could demonstrate many distinctive physics, for example, ferroelectric control of itinerant AFM phase and the sign of quantized anomalous Hall plateau. The explored phenomena and mechanism would not only enrich the research in 2D ferroelectricity and topological magnets, but also guide the design of low-consumption high-speed quantum devices.

Physically, the symmetry of a crystalline material governs its functional properties [1-3]. By breaking the inversion (*P*) symmetry, spontaneous electric polarization could be obtained. If the polarization is switchable and responds nonlinearly to external electric field, the exotic ferroelectricity will be achieved [4]. Depending on the direction of the electric polarization, the reported ferroelectricity in two-dimensional lattices normally can be classified into out-of-plane and in-plane ferroelectricity [5,6]. Going beyond these two classes, recent discovery of 2D intercorrelated ferroelectrics (InFE) receives special attention as it suggests new device paradigms [7,8]. Different from conventional ferroelectrics, InFE requires the coexistence and strong coupling of in-plane (IP) and out-of-plane (OOP) polarizations, and thus is highly desirable but rare in natural [7,9,10]. Analogues to the case of broken *P* symmetry, the broken of time-reversal (*T*) symmetry leads to spontaneous spin polarization. Such spin polarization could induce many exotic phenomena, such as anti-ferromagnetism (AFM) and quantum anomalous Hall (QAH) effect, which ignites profound scientific and technological impacts [11-14]. Particularly, with the recent discovery of 2D magnetism in $CrI_3$ [15], $Cr_2Ge_2Te_6$ [16] and $MnBi_2Te_4$[13,17], rapid development has been made in AFM and QAH effect.

Compared with breaking *P* or *T* symmetry independently, undoubtedly, the simultaneous breaking of *P* and *T* symmetries is more interesting as it might couple these characteristic features and thus demonstrate novel exotic phenomena [18-20]. However, the simultaneously occurrences of multi-parameters are rather scarce. For example, magnetic ferroelectrics is hindered by the inherent exclusion between ferroelectricity and magnetism [21]. AFM QAHE is limited by the combined effect of magnetization and SOC, in which ferromagnetism is deemed as one essential precondition for QAHE in most cases [13,14,22]. No example has been proposed for FE QAHE, despite there is no inherent exclusion between electric polarization and nontrivial band topology. Actually, so far, the coupling among InFE, AFM and QAH effect has not been reported.

In this letter, by taking four-layer AFM $MnBi_2Te_4$ as an example, we demonstrate a novel mechanism for realizing of the coupling among InFE, AFM and QAH effect in 2D van der Waals (vdW) materials. Using first-principles calculations, we show that four-layer $MnBi_2Te_4$ (FLMBT) exhibits AFM, InFE and QAHE simultaneously through interlayer sliding. More remarkably, the AFM, FE and QAH orders are bound together strongly, which enables the scenario of non-volatile ferroelectricity control of itinerant AFM and the sign of quantized anomalous Hall plateau (conductivity). Our findings reveal the possibility of accessing experiment-friendly stoichiometric systems spanning correlated FE, AFM and QAH orders, which are important to future exotic physics observations and innovative technologies explorations.

Our calculations are performed based on density functional theory as implemented in the Vienna *Ab Initio* Simulation Package (VASP) [23]. The generalized gradient approximation (GGA) as formulated by the scheme of Perdew, Burke, and Ernzerhof (PBE) is used to estimate the exchange-correlation [24]. Following previous investigations, GGA + $U$ method with $U - J = 5.34$ eV is applied to Mn 3*d*-states to reduce the self-interaction error, and the dispersion-corrected DFT-D3 method is introduced to take interlayer van der Waals interaction into account [25,26]. The vacuum space is set to be larger than 18 Å to eliminate the interactions between periodic images. Convergence threshold of energy and forces are $10^{-4}$ eV and 0.02 eVÅ$^{-1}$, respectively, at the kinetic cutoff energy of 400 eV. A Monkhorst-Pack *k*-point sampling of $9 \times 9 \times 1$ is utilized for the Brillouin zone integration. Wannier charge center (WCC), quantum anomalous Hall conductance and edge-states calculations are performed based on maximum localized Wannier functions by Wannier90 and WannierTools packages [27,28]. The ferroelectric switching pathway is calculated via nudged elastic band (NEB) method [29].

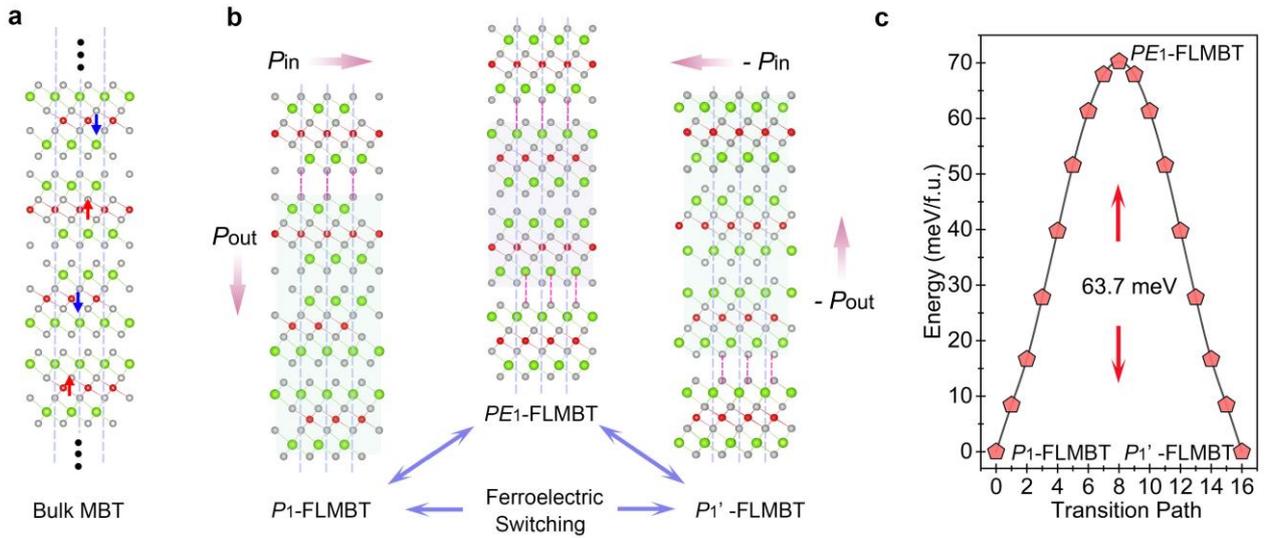

**FIG. 1.** (a) Crystal structure of bulk MnBi$_2$Te$_4$. (b) Atomic structures of $P_1$-FLMBT and $P_1'$-FLMBT and the corresponding FE switching mechanism. (c) Pathway of switching between $P_1$-FLMBT and $P_1'$-FLMBT by nudged elastic band calculation. The IP and OOP polarizations are marked with pink arrows.

Note that it is quite difficult to find intrinsic host exhibiting FE, AFM and QAHE characteristics, we herein introduce the scenario of vdW stacking sequence control. The successful experimental trials in discovering 2D sliding ferroelectrics driving by interlayer stacking, including WTe$_2$ [30], BN [31,32], 1T'-ReS$_2$ [33,34] and MoS$_2$ layers [35], confirming the solid foundation of universal existence of the interlayer ferroelectricity in 2D layers proposed by Wu *et al* [36]. This sliding ferroelectricity originates from the non-centrosymmetric coordination, which is switchable by vertical electric field induced in-plane interlayer sliding with ultralow energy barrier. Considering diverse physical properties of 2D monolayers, such as excitons, nontrivial topology, magnetism and valley pseudospin, potential coupling between sliding ferroelectricity and fertile physical phenomena can be uncovered [9,37-39]. Recently, a new type of magnetism-dependent topological materials $MnBi_2Te_4$ has attracted great attention [13,40-42]. In the following, we present the coexistence and tight coupling relationship between sliding FE, AFM and QAHE in multilayer $MnBi_2Te_4$ model system.

Ferroelectricity by definition is a class of condensed matter with spontaneous electric polarization that is switchable by an applied electric field. Fig. 1a plots the crystal structure of bulk $MnBi_2Te_4$, in which each single layer (SL) consists of seven atomic layers alternate in a sequence Te-Bi-Te-Mn-Te-Bi-Te, while the neighboring SLs connect via vdW interaction with AB stacking. Moreover, intralayer Mn spins couple ferromagnetically, but the adjacent Mn layers couple antiparallel to each other. Our calculated lattice constant, magnetism and topology agree well with previous studies [26,43]. Due to the structural centrosymmetry, electric polarization is not allowed in both SL and bulk $MnBi_2Te_4$. Fortunately, ferroelectricity could be expected in vdW multilayer, once the centrosymmetry or out-of-plane mirror symmetry broken induced electric polarization is switchable electrically. Thus, introducing ferroelectricity by stacking sequence control via interlayer sliding is highly plausible. Bilayer $MnBi_2Te_4$ is excluded first for its trivial band topology and preserved P symmetry against interlayer sliding (Supplemental Material I). Moreover, to realize the equivalent ferroelectric states, AFM layers is preferred [36,38]. Therefore, ferrimagnetic trilayer $MnBi_2Te_4$ is also not under our consideration and naturally four-layer $MnBi_2Te_4$ is chosen in the next discussions.

FLMBT hosts no FE due to the preserved P symmetry. We thus slid uppermost and lowest SL along opposite directions, respectively, which are denoted as $P_1$-FLMBT and $P_1{'}$-FLMBT (Fig. 1b). This kind of layer sliding breaks structural *P* symmetry, giving rise to internal charge redistribution within the layers. Moreover, for $P_1$-FLMBT shown in left panel of Fig.1b, it is clear that central bilayer $MnBi_2Te_4$ exhibits inversion symmetry. However, the inversion center of central bilayer is closer to bottom $MnBi_2Te_4$ layer than that of upmost layer in both OOP and IP directions, where the unequal environments between top and bottom SLs result in a spontaneous electric polarization of $8.16 \times 10^{10}$ e/cm$^2$ and $1.35 \times 10^{10}$ e/cm$^2$, respectively. $P_1$-FLMBT and $P_1{'}$-FLMBT harbor opposite electric polarizations in both IP and OOP directions, and are exchangeable upon an 180º rotation operation with respect to the orthogonal direction of IP and OOP polarizations. Obviously, similar to experimental observations, these two configurations are switchable via interlayer sliding and can be deemed as two ferroelectric states. Both IP and OOP ferroelectricity flips accompany with central bilayer translation relative to outmost layers, suggesting the appearance of InFE with strongly coupled IP and OOP ferroelectricity.

To provide insights in understanding the ferroelectricity emanating from the vdW sliding, planner averaged electrostatic potentials is further analyzed. The inequivalent atomic distributions give rise to non-overlapping positive and negative charge centers, as reflected by the discontinuity between the vacuum levels of the top and bottom layers in $P_1$-FLMBT and $P_2$-FLMBT, which yields out-of-plane polarizations (Supplemental Material II). To further elucidate the feasibility of anticipated ferroelectricity, the electric field induced polarity reversal is characterized kinetically. As shown in Fig. 1c, minimum energy path for polarization flip goes through a centrosymmetric configuration $PE_1$-FLMBT. And the corresponding energy barrier equals 63.7 meV per unit cell, which is comparable to other sliding ferroelectrics [9,34,44], but much smaller than traditional ferroelectric systems with requisite ion displacement [8,45,46]. Such an ultralow switching barrier not only indicate its high experimental feasibility, but also enables high-speed data writing with low energy cost. Similarly, there is another set of ferroelectric configurations, $P_2$-FLMBT and $P_2'$-FLMBT, which are described in Supplemental Material III. To simplify, we mainly focus on $P_1$-FLMBT and $P_1'$-FLMBT in the following.

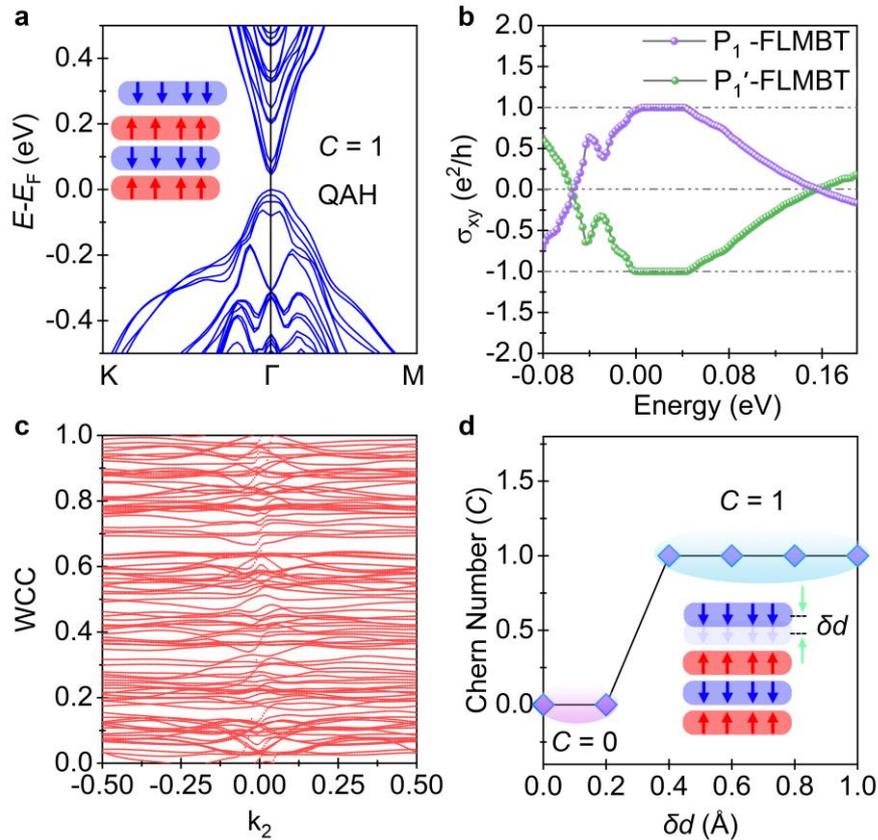

**FIG. 2.** (a) Band structure of $P_1$-FLMBT with SOC. (b) The anomalous Hall conductivity ($\sigma_{xy}$) for $P_1$-FLMBT and $P_1'$-FLMBT near the Fermi level. (c) WCC with close momentum surface in $P_1$-FLMBT. (d) Evolution of the Chern number ($C$) as a function of interlayer distance increasing ($\delta d$) between surface and remaining trilayer in pristine FLMBT.

Having verified the sliding ferroelectricity, we next study the electronic properties of ferroelectric FLMBT in detail. Supplemental Material IV shows the band structures of $P_1$-FLMBT without spin-orbit coupling (SOC). It is obvious that the appearance of ferroelectricity splits the degenerate bands, and the global gap is calculated to be 608 meV. While for $P_1'$-FLMBT with reversed electric polarization, the microscopic magnetic spin distribution is reversed upon ferroelectric switching, and in turn spin bands exchange with respect to $P_1$-FLMBT with retained gap size. Thus, basically, the ferroelectric switching causes the swapping of the spin-up and spin-down channels, endowing the electrical writing that can be read efficiently by the magnetic carrier type of the sample. When SOC is considered, the CBM and VBM of $P_1$-FLMBT get closer and the band gap is reduced to 45 meV, as shown in Fig. 2a, which is 23 meV smaller than pristine FLMBT in our numerical calculations. Comparing with pristine FLMBT, the symmetry combines spatial inversion and time reversal as well as the band structure are modified strongly upon interlayer sliding, we expect it may induce a distinct topological phase.

To determine the band topology, the **k**-resolved Berry curvatures of $P_1$-FLMBT and $P_1'$-FLMBT are obtained by analyzing the Bloch wave functions [47],

$$\Omega(\boldsymbol{k}) = \sum_n f_n \Omega_n(\boldsymbol{k})$$

$$\Omega_n(\boldsymbol{k}) = -\sum_{n' \neq n} \frac{2Im\langle\psi_{n\boldsymbol{k}}|v_x|\psi_{n'\boldsymbol{k}}\rangle\langle\psi_{n'\boldsymbol{k}}|v_y|\psi_{n\boldsymbol{k}}\rangle}{(\omega_{n'} - \omega_n)^2}$$

Here $n$ denotes band index, $f_n$ is the Fermi-Dirac distribution function, $\omega_n$ is the eigenvalue of $|\psi_{n\boldsymbol{k}}\rangle$, $v_x$ and $v_y$ are the velocity operators. The results of summation run over all the occupied bands are shown Supplemental Material IV, it is found that nonzero $\Omega(\boldsymbol{k})$ distributed mainly around $\Gamma$ point. The only difference between $P_1$-FLMBT and $P_1'$-FLMBT is that, the sign of $\Omega(\boldsymbol{k})$ is opposite to each other at every $\boldsymbol{k}$ points. Surprisingly, further integrating of $\Omega(\boldsymbol{k})$ over first BZ by

$$C = \frac{1}{2\pi} \int_{BZ} \Omega(\boldsymbol{k}) d^2\boldsymbol{k}$$

The nonzero integer Chern numbers of $C$ = 1 and -1 are obtained respectively for $P_1$-FLMBT and $P_1'$-

FLMBT. This is also consistent with the calculated flat plateau of anomalous Hall conductance ($\sigma_{xy}$) corresponding to $1 \times e^2/h$ and $-1 \times e^2/h$ near the Fermi level, as displayed in Fig. 2b, which strongly validates QAH effect in ferroelectric FLMBT. Additionally, other two prominent features essential for QAH effect, i.e. Wannier charge center (WCC) and one chiral edge state, are also obtained and further manifest the nontrivial properties, as can be seen in Fig. 2c. As a result, all these results manifest the ascription of $P_1$-FLMBT as a ferroelectric QAH insulator. Besides, its nontrivial band gap is much larger than room temperature thermal energy ~26 meV, ensuring ferroelectric controllable QAH effect at high working temperature.

At odds with previous observations, where even-layer AFM MBT are believed to be axion insulators [26,40,48]. It is really interesting to find ferroelectric FLMBTs ($P_1$-FLMBT and $P_2$-FLMBT) are AFM QAH insulators. Here, we ascribe the quantum phase transitions to sliding induced interlayer coupling variation. The difference between pristine and ferroelectric FLMBTs are distinguishable from interlayer distance between topmost SL and the remaining trilayer, which are calculated to be 2.81, 3.85 and 3.17 Å for pristine, $P_1$-FLMBT and $P_2$-FLMBT, respectively. Smallest interlayer distance in pristine FLMBT gives strongest interlayer coupling, which exhibits different quantum phase compared with $P_1$-FLMBT and $P_2$-FLMBT. This is validated from quantum phase transition in pristine FLMBT by changing this factor. Decreasing the interlayer coupling between topmost layer and the remaining trilayer, by increasing interlayer distance between topmost and remaining trilayer ($\delta d$) as shown in Fig. 2d, indeed triggers the transition from trivial insulator to QAH insulator. Since the decreased coupling in $P_1$-FLMBT and $P_2$-FLMBT, it can be considered that topmost SL serves as a "trivial substrate", which is irrelevant to the QAH effect of remaining trilayer. These analyses unambiguously demonstrate the reason for the appearance of QAH phases in InFE AFM FLMBTs. We provide a physical picture where the QAH phase is driven by interlayer coupling in even layer AFM MBT. In addition, we suggest experimentalists to explain or toggle distinct topological phases via this interlayer coupling mechanism.

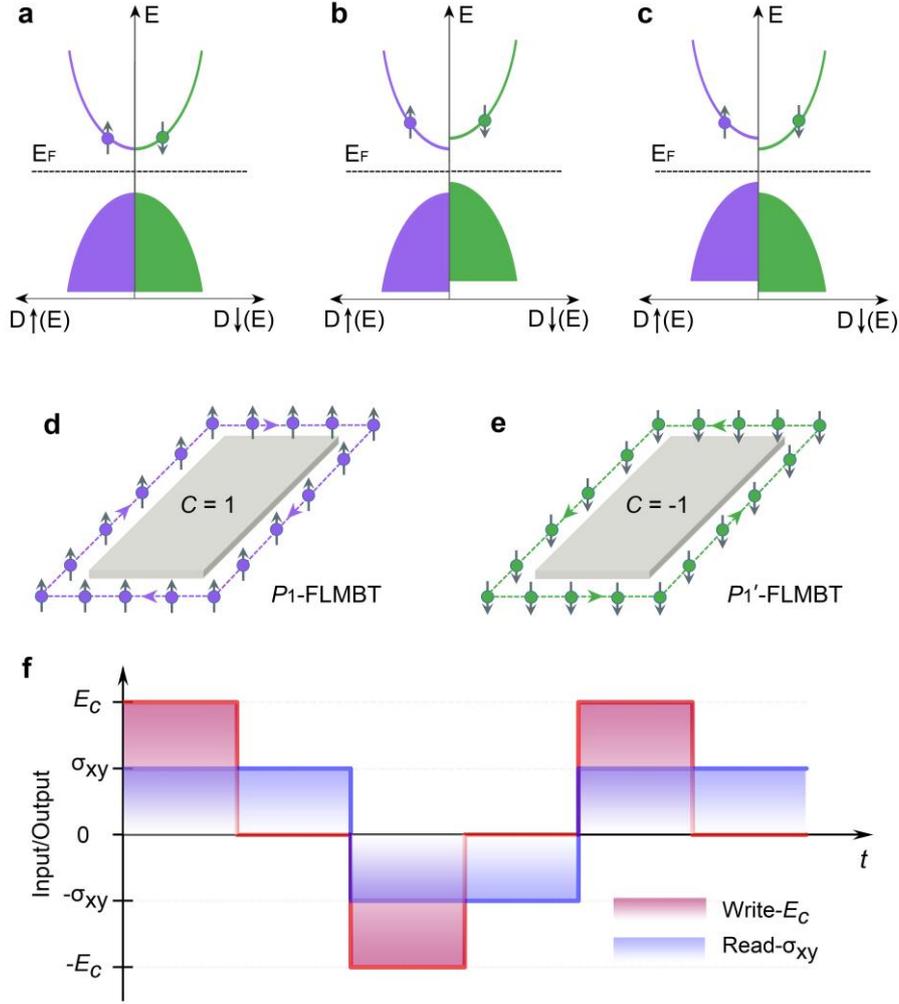

**FIG. 3.** Schematic spin-resolved density of states of FLMBT: (a) without out-of-plane electric polarization; (b) and (c) with opposite out-of-plane electric polarization. (d, e) Illustration of chiral edge current propagation for the two ferroelectric phases of FLMBT. (f) Schematic operation for the devices based on ferroelectric AFM QAH insulator FLMBT. Coercive field $E_c$ is the magnitude of the required electric field for reversing the polarization. $\sigma_{xy}$ equals $1 \times e^2/h$ and $-1 \times e^2/h$, respectively.

As the coexistence of InFE, AFM and QAH orders are demonstrated, we next move to reveal the advances and potential applications based on their coupling. Set $P_1$-FLMBT as an example, it is interesting to aware that $P_1$-FLMBT and $P_1'$-FLMBT possess opposite QAH conductance at given chemical potential. Since $P_1$-FLMBT and $P_1'$-FLMBT are two equivalent ferroelectric states, expected electric and spin polarization as well as the AH conductance are therefore electrically controllable with nonvolatile and low energy consumption characteristics, as depicted in Fig.3 a-e. To date, tunable AH conductance can only be achieved by reversing the magnetization directly or through spin-orbit torque [49,50], which are cumbersome and volatile procedures. In contrast, circumstance reported here gives purely nonvolatile electrical way beyond previous approaches due to the association between ferroelectricity and internal magnetization direction. For data storage, since ferroelectric states are controllable by both in-plane and out-of-plane voltages, encoding the information as the "write-in" processes and the detection of AH conductance as the "read-out" processes. As shown in Fig. 3f, if two ferroelectric states switched by electric field $E_c$ and -$E_c$, in principle, the square output AH signal converts between $1 \times e^2/h$ and $-1 \times e^2/h$ when suffering in-plane/out-of-plane pulse voltage. In addition, the evolution of AH conductance exhibit hysteresis feature as a function of electric field, owing to the nonvolatility of ferroelectricity. Removing external electrically input, output AH signal survives, paving the way to the absolutely new technology breakthroughs concerning information technology and AH conductance-based electronics and spintronics.

It should be noted that, beyond the FLMBT, the discussed physics and proposed device prototype in this work could provide a general mechanism for other sliding manipulated even layer $MnBi_2Te_4$ thicker than four layers. Moreover, we also wish to emphasize that the intimate relationship among ferroelectricity, AFM and QAH orders produced by van der Waals sliding are not limited to FLMBT, but ubiquitous in other $MnBi_2Te_4$-family materials with interlayer AFM and "Bi/Sb-Te … Te-Sb/Bi" interfaces [40]. Furthermore, $MnBi_2Te_4$ shows ordered structures with large band gap, inherent to the stoichiometric material, which protects them against drawbacks such as disorder, inhomogeneity and disturbance of trivial bands. In particular, it is an easily accessible material in current experiments, significantly facilitates the research of InFE AFM QAH insulator in condensed matter fields. We thus believe that our findings will stimulate intensive studies of antiferromagnets topological insulators, especially $MnBi_2Te_4$, as prospective materials for ferroelectric AFM quantum spintronics. These results are awaiting experimental efforts.

In summary, we report the theoretical evidence of intercorrelated ferroelectric AFM QAH insulators in single-phase prototype materials, experimental accessible FLMBT, by van der Waals sliding. The intercorrelated ferroelectricity and QAH phase originated from the sliding modification of interface charge redistribution and interlayer coupling. We also demonstrate that it allows nonvolatile control of charge, spin polarization and AH signal by purely ferroelectricity. These findings open a new direction for exploring the interesting correlation between ferroelectricity and magnetic topological insulators, as well as the potential applications of crossing quantum states in low-power-consumption controllable electronics.

## ACKNOWLEDGEMENTS

This work is supported by the DFG Research and Training Group RTG 2247 "Quantum Mechanical Materials Modelling", Qingdao Postdoctoral Science Foundation (No. QDBSH20220202095) and